\documentclass{article}
\usepackage{ijcai20}

\usepackage{times}
\usepackage{soul}
\usepackage{url}
\usepackage[hidelinks]{hyperref}
\usepackage[small]{caption}
\usepackage{graphicx}
\usepackage{amsmath}
\usepackage{amsthm}
\usepackage{booktabs}
\usepackage{algorithm}
\usepackage{algorithmic}
\usepackage{amssymb}
\title{Random-walk Based Generative Model for Classifying Document Networks}

\author{
    Takafumi J. Suzuki
    \affiliations
    Communication Technology Laboratory, Research {and} Technology Group, Fuji Xerox Co., Ltd.
    6-1 Minatomirai, Nishi-ku, Yokohama, Kanagawa, 220-8668, Japan
    \emails
    suzuki.taka-fumi@fujixerox.co.jp
}

\begin{document}

\maketitle

\begin{abstract}
Document networks are found in various collections of real-world data, such as citation networks, hyperlinked web pages, and online social networks.
A large number of generative models have been proposed because they offer intuitive and useful pictures for analyzing document networks.
Prominent examples are relational topic models, where documents are linked according to their topic similarities.
However, existing generative models do not make full use of network structures because they are largely dependent on topic modeling of documents.
In particular, centrality of graph nodes is missing in generative processes of previous models.
In this paper, we propose a novel generative model for document networks by introducing random walkers on networks to integrate the node centrality into link generation processes.
The developed method is evaluated in semi-supervised classification tasks with real-world citation networks.
We show that the proposed model outperforms existing probabilistic approaches especially in detecting communities in connected networks.
\end{abstract}

\section{Introduction}

Graph representation is one of the most fundamental data structures in computer science, and describes various types of real-world data such as social networks, relational databases, and the world-wide web.
Uncovering clusters, which are also referred as {\it communities} in network science, is an essential step to clarify intrinsic natures of concerned data.
There has been a lot of works on community detection algorithms including supervised and unsupervised approaches~\cite{FORTUNATO20161}.
Among the various methods, Bayesian approaches to graph generation processes have been intensively studied over many years because they give us fundamental insights on hidden patterns of networks.

Usually, real-world networks are composed of not only link structures between nodes but also rich information on their constituents.
Citation networks belong to an important class of such networks, where each article contains informative sets of words to topology of citation links.
Since topic analysis of citation networks enables us to predict latent citation links, a lot of generative models have been proposed to utilize the complementary information from graph and contents.
Relational topic model (RTM)~\cite{pmlr-v5-chang09a} is an extension of the latent Dirichlet allocation (LDA)~\cite{Blei:2003:LDA:944919.944937} to predicting citation links based on topic similarities.
RTM has been further extend to incorporate various aspects of underlying topics.
For instance, generalized RTM (gRTM)~\cite{Chen2015} takes into account inter-topic similarities.
Constrained RTM~\cite{TERRAGNI2020581} reflects prior knowledge on documents to link generation processes.
Besides LDA, traditional generative models including probabilistic latent semantic analysis (PLSA) and stochastic block model (SBM) have been flexibly employed to describe citation networks~\cite{cohn2001missing,Erosheva5220,nallapati2008link,Liu2009,yang-etal-2016-discriminative}.

In order to develop generative models which jointly describe networks and texts, it is crucial to capture intrinsic natures of both data types.
From this perspective, existing models do not seem to fully utilize topological structures because they are basically developed upon topic modeling of documents.
RTM and its variants, for instance, are only concerned with topic similarities between local pairs of documents, while non-local coherence between multiples of documents are not considered.
Meanwhile, it is widely known in network science that random walks can capture global information of networks.
One of the most successful examples is PageRank algorithm~\cite{brin1998anatomy}, where centrality of nodes is evaluated with eigenvectors of modified transition matrices.
Recently, Okamoto and Qiu~\shortcite{mdmc_original,okamoto2019modular} have proposed a novel generative model called modular decomposition of Markov chain (MDMC) for network clustering.
The key idea of MDMC is to introduce random walkers on networks to utilize global link structures for detecting communities.
Predictive performance of MDMC has been further elaborated with Gibbs sampling algorithms, and becomes competitive with other probabilistic community-detection approaches~\cite{suzuki2019bayesian}.
Hence, it is fruitful to unify MDMC and basic topic models to simultaneously leverage the complementary information from latent topics and communities.







In this paper, we develop a novel generative model named topic MDMC (TMDMC), where a random-walk method is augmented with additional textual information to improve detectability of community structures.
We apply TMDMC to transductive classification tasks with real-world citation networks, and show that TMDMC outperforms other probabilistic approaches especially in detecting community in connected networks.
The remainder of the paper is organized as follows:
In Section \ref{sec: related works}, we review previous models for citation networks.
In Section \ref{sec: preliminaries}, we outline MDMC to provide a basis for random-walk approaches.
We combine MDMC with topic modeling and propose our model in Section \ref{sec: Bayesian Formulation of TMDMC}.
TMDMC is evaluated in semi-supervised node classification for benchmark citation networks in Section \ref{sec: experiments}.
Conclusions and promising future works are stated in Section \ref{sec: discussion and conclusion}.


\section{Related Works}
\label{sec: related works}

Joint modeling of network structures and document contents has been intensively studied to perform traditional tasks such as node classification and link prediction.
Broadly speaking, these approaches fall into two major categories: probabilistic and deterministic approaches.


The proposed method shares some common features with previous probabilistic approaches.
RTM~\cite{pmlr-v5-chang09a} is a hierarchical probabilistic model associating links between documents with their topic similarities:
Links are considered to be more often formed between documents with similar topic distributions, which are estimated by LDA~\cite{Blei:2003:LDA:944919.944937} in advance.
While original RTM allows interactions only within the same topics, gRTM~\cite{Chen2015} incorporates inter-topic correlations with weighted matrices in a latent topic space.
Imbalance issues between presence and absence of links in observed graphs are also alleviated in gRTM with regularized Bayesian inference techniques.
Bai {\it et al.} \shortcite{Bai2018} have employed neural architectures to overcome limited expressivity of RTM.
Very recently, RTM has been extend to a semi-supervised model to incorporate prior knowledge on must-link and cannot-link constraints~\cite{TERRAGNI2020581}.
While link generations are considered as downstream tasks in RTMs, some approaches consider them as parallel or upstream processes.
Cohn and Hofmann~\shortcite{cohn2001missing} have used PLSA as a core building block to perform a simultaneous decomposition of texts and graphs.
PLSA has been replaced by LDA in Erosheva {\it et al.}~\shortcite{Erosheva5220}, whose graphical model is closely related to our proposed model.
SBM, which 
is considered as one of the most famous generative models for community detection,
has been also integrated with RTM to model document networks in LBH-RTM~\cite{yang-etal-2016-discriminative}.


In spite of intensive studies, most of the LDA-based models do not fully utilize network structures because they consider link generation processes as downstream tasks of topic modeling.
An important exception is LBH-RTM~\cite{yang-etal-2016-discriminative}, which associates link generation processes with latent communities by weighted SBM.
However, it may discard crucial information from node centrality because weighted SBM cannot resolve nodes in communities.
On the other hand, random walks, which are integrated in our proposed model, can quantify node centrality, and provide the rich information for various downstream tasks.

Another successful approach for analyzing document networks is to deterministically embed graph nodes into low-dimensional feature vectors, which are later used in node classification or link prediction.
Traditional methods, such as label propagation~\cite{zhu2003semi}, manifold regularization~\cite{belkin2006manifold}, use graph Laplacian as regularization terms in the corresponding loss functions.
Neural architectures have also been frequently employed to learn node vectors.
Semi-supervised embedding~\cite{weston2012deep} imposes regularization terms in deep architectures to learn graph structures.
Deep Walk~\cite{perozzi2014deepwalk} embeds nodes into a low dimensional space with node sequences obtained by random walkers.
Planetoid~\cite{Yang:2016:RSL:3045390.3045396} has extended Deep Walk to jointly embed node features and link structures with Skipgram models.
Recently, limited expressivity of traditional models has been significantly relaxed by graph convolutional networks (GCNs)~\cite{Kipf2017}, whose hidden layers are used as node embedding vectors.
In spite of the performance, however, vast amounts of model parameters must be optimized in GCNs, which require techniques such as renormalization tricks~\cite{Kipf2017} and kernel smoothing~\cite{xu2019graph}.
Establishing efficient and powerful schemes in GCNs is a challenging on-going issue.

\section{Preliminaries}
\label{sec: preliminaries}

In this section, we outline generative processes of MDMC \cite{mdmc_original,okamoto2019modular,suzuki2019bayesian} to lay the base of our new model.
The key idea of MDMC is to introduce and observe Markovian dynamics of random walkers who travel around networks.
The network structures are characterized by transition matrix $T_{mn}$, which satisfies $\sum^{N}_{m=1}T_{mn}=1$.
Given transition probabilities $p^{(t)}$ of the agent at time $t$, those at the next time are designed to be equal to $T p^{(t)}$ in terms of expectation values.
Each link in observed graphs can be encoded into $N$-dimensional vectors $\tau^{(t)}_{l}$, where $\tau^{(t)}_{ln}=1$ if link $l$ contains node $n$ and otherwise $0$.
Parameters in MDMC are optimized in an unsupervised manner to reconstruct observed graph $\tau^{(t)}$.
MDMC models generative processes of links by combining agent probability distributions $p^{(t)}$ and latent community assignments $z^{(t)}$.

MDMC instantiates the aforementioned ideas with following generating processes at time $t$.
\begin{enumerate}
\item For each community $k=1,2,\cdots,K$:
\begin{enumerate}
\item
Draw probabilities
$p^{(t)}(:|k) \sim \text{Dir}(\alpha^{(t)}_{:k})$
with
$\alpha^{(t)}_{nk} = \alpha^{(t)}_{k} \sum^{N}_{m=1} T_{nm}p^{(t-1)}(m|k)$.
\end{enumerate}
\item For each link $l=1,2,\cdots,L$:
\begin{enumerate}
\item
Draw community distributions
$ \pi^{(t)}_{l} \sim  \text{Dir}(\eta^{(t)})$.
\item
Draw community assignment
$ z^{(t)}_{l} \sim \text{Mult}\left( \pi^{(t)}_{l} \right)$.
\item
  Draw link data $\tau^{(t)}_{l}
  \sim \text{Mult}\left( p^{(t)}(:|z^{(t)}_{lk}=1)\right)$.
\end{enumerate}
\end{enumerate}
Here, $\text{Dir}(\cdot)$ and $\text{Mult}(\cdot)$ denote the Dirichlet and multinomial distributions, respectively.

Although generative processes in MDMC closely resemble those in LDA, there are several crucial differences for analyzing network structures.
The most important point is that prior distribution of $p^{(t)}$ is dependent on transition matrix $T$ and previous distribution $p^{(t-1)}$.
This modeling instantiates the Markovian dynamics of random walkers who capture global network structures.
The second point is that generation probability of link $l$ connecting nodes $m$ and $n$ is proportional to $p^{(t)}(m|k)p^{(t)}(n|k)$ with presumed community $k$.
An intuition behind this modeling is that links are more often generated between central nodes within a community.


\section{Augmenting MDMC with Text information}
\label{sec: Bayesian Formulation of TMDMC}

In this section, we detail our proposed model, which we call TMDMC.
Subsection \ref{subsec: Generative processes of TMDMC} describes how to represent textual information and presents generative processes of TMDMC.
Parameter inference steps and procedures of implementing TMDMC are discussed in Subsection \ref{subsec: Parameter Inference}.
TMDMC is extended to a semi-supervised model for node classification tasks in Subsection \ref{subsec: Semi-supervised}.

\subsection{Topic MDMC}
\label{subsec: Generative processes of TMDMC}

In order to model generative processes of documents, we consider documents as sets of words, i.e. bag of words (BoW) representation.
We will incorporate the BoW-represented documents into an MDMC scheme with following strategies:
(1) We introduce ``linked documents'' $w^{(t)}_{li}$ for link $l$ by combining BoW-representation of endpoint documents.
(2) Topic $y^{(t)}_{li}$ is assigned to each word $i$ in linked documents with topic distribution $\pi^{(t)}_{l}$, which are also used to predict community assignments of the concerned link.
(3) With topic $y^{(t)}_{li}$ assigned in the previous step, we fit the observed word $w^{(t)}_{li}$ with topic-specific word distribution $\phi^{(t)}$.
This modeling is motivated by the observation that documents linked in networks should share similar topic distributions, which are also common to communities of links.
Consistency between link communities and word topics are guaranteed through the common distribution $\pi^{(t)}_{l}$.

\begin{figure}[t]
\begin{center}
\includegraphics[width=70mm]{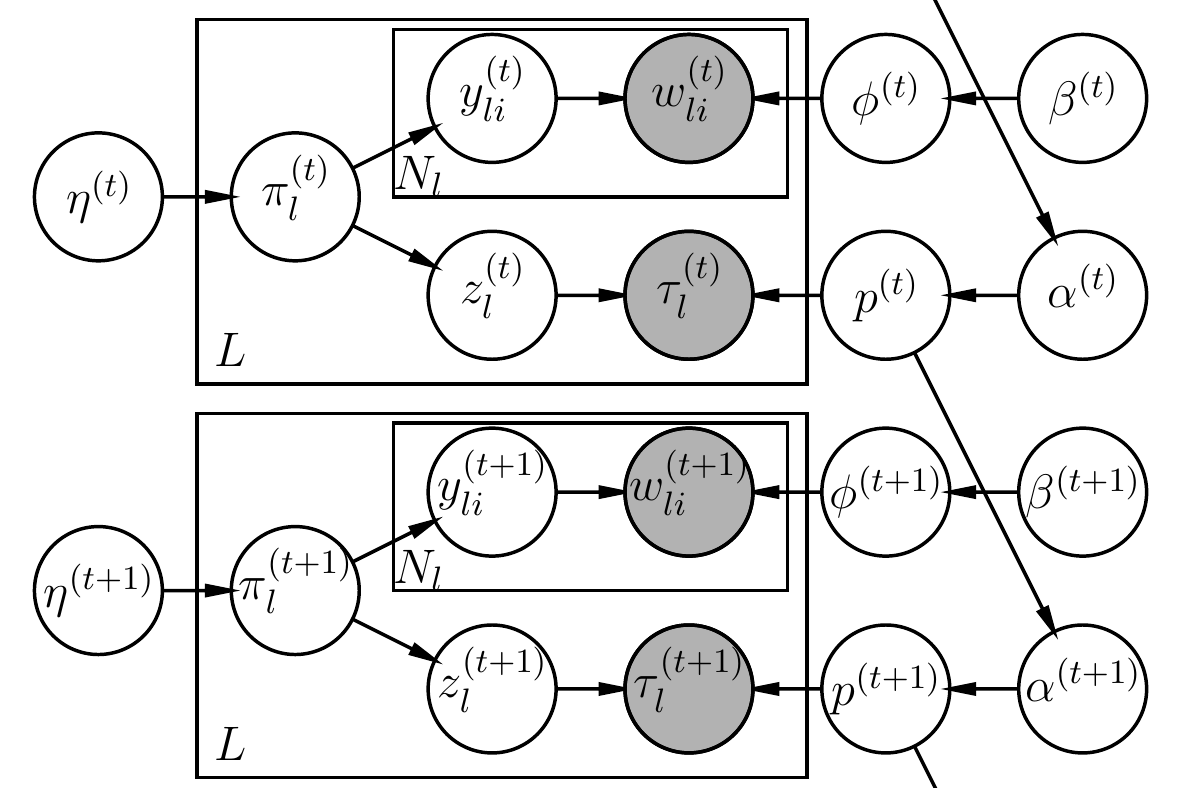}
\end{center}
\caption{Graphical model of TMDMC.}
\label{fig: graphical representation of TMDMC}
\end{figure}

TMDMC consists of the following additional generative processes at time $t$ on top of the original MDMC model.
\begin{enumerate}
\item For each community $k=1,2,\cdots,K$:
\begin{enumerate}
\item[(b)]
Draw word distributions
$\phi^{(t)}(:|k) \sim \text{Dir}(\beta^{(t)}_{:k})$.
\end{enumerate}
\item For each link $l=1,2,\cdots,L$:
\begin{enumerate}
\item[(d)] For each word $i=1,2,\cdots,N_{l}$:
\begin{enumerate}
\item Draw topic vectors
$ y^{(t)}_{li} \sim \text{Mult}\left( \pi^{(t)}_{l} \right)$.
\item Draw words 
  $w^{(t)}_{di}
  \sim \text{Mult}\left( \phi^{(t)}(:|y^{(t)}_{lik}=1)\right) $.
\end{enumerate}
\end{enumerate}
\end{enumerate}
The graphical model representation of TMDMC is shown in Figure~\ref{fig: graphical representation of TMDMC}.
The model is composed of time-series blocks, where upper and lower branches within each block represent LDA and original MDMC, respectively.
The inter-block links are introduced by the Markovian dynamics of random walkers inherited from original MDMC.

\subsection{Parameter Inference}
\label{subsec: Parameter Inference}

\paragraph{Approximation.}

While TMDMC theoretically consists of infinite numbers of sequential blocks which are dependent on agent probability distributions at previous time steps, it is hard to simultaneously optimize the whole parameters.
In order to make the model tractable, we approximate the hyper-parameters $\alpha^{(t)}_{nk}$ to their expectation values~\cite{suzuki2019bayesian}.
This approximation cuts dependency between different blocks in parameter inference steps because it is sufficient to predict the expectation value of $p^{(t)}$ to infer the parameters at next Markov time.


\paragraph{Joint probability distribution}
Thanks to the approximation described above, it is allowed to separately optimize the parameters at each Markov time.
We achieve the joint probability distribution at time $t$ by marginalizing $p^{(t)}$, $\pi^{(t)}$, and $\phi^{(t)}$ as
\begin{align}
  & P(\tau^{(t)},w^{(t)},z^{(t)},y^{(t)}) \nonumber \\
  & \propto
  \frac{\Gamma\left( \sum^{K}_{k=1} \eta^{(t)}_{k} \right)}{\prod^{K}_{k=1} \Gamma\left(\eta^{(t)}_{k}\right)}
  \frac{\prod^{K}_{k=1}\Gamma\left( \eta^{(t)}_{k}+ \sum^{L}_{l=1}z^{(t)}_{lk}\right)}{\Gamma\left(\sum^{K}_{k=1} \left( \eta^{(t)}_{k}+ \sum^{L}_{l=1}z^{(t)}_{lk}\right)\right)}
  \nonumber \\
  & \hspace{10pt} \times
  \prod^{K}_{k=1}
  \left[
    \frac{\Gamma\left( \sum^{N}_{n=1} \alpha^{(t)}_{nk} \right)}{\prod^{N}_{n=1} \Gamma(\alpha^{(t)}_{nk})}
    \frac{\prod^{N}_{n=1}\Gamma\left( \alpha^{(t)}_{nk}+ (\tau z)^{(t)}_{nk}\right)}{\Gamma\left(\sum^{N}_{n=1} \left( \alpha^{(t)}_{nk}+ (\tau z)^{(t)}_{nk}\right)\right)}
    \right.
  \nonumber \\
  \label{eq: Marginalized likelihood function TMDMC}
  & \hspace{30pt}
  \times 
  \left.
    \frac{\Gamma\left( \sum^{V}_{w=1} \beta^{(t)}_{wk} \right)}{\prod^{V}_{w=1} \Gamma(\beta^{(t)}_{wk})}
    \frac{\prod^{V}_{w=1}\Gamma\left( \beta^{(t)}_{wk}+ Y^{(t)}_{wk}\right)}{\Gamma\left(\sum^{V}_{w=1} \left( \beta^{(t)}_{wk}+ Y^{(t)}_{wk}\right)\right)}
    \right],  
\end{align}
with
$(\tau z)^{(t)}_{nk} = \sum^{L}_{l=1} \tau^{(t)}_{ln}z^{(t)}_{lk}$
and
$Y^{(t)}_{wk}=\sum_{(l,i)}\delta_{w,w_{li}} y^{(t)}_{lik}$,
where
$\delta_{ab}$ denotes Kronecker delta.

\paragraph{Collapsed Gibbs Sampling.}
With the aid of Eq.~(\ref{eq: Marginalized likelihood function TMDMC}), community $z^{(t)}$ and topic $y^{(t)}$ assignments are sampled as
\begin{align}
  \label{eq: z Collapsed Gibbs Sampling TMDMC}
  &P(z^{(t)}_{lk}=1|\tau^{(t)},w^{(t)},z^{(t)}_{\backslash l},y^{(t)}) \nonumber \\
  &\propto 
  \left(\eta^{(t)}_{k}+Y^{(t)}_{l.k}\right)
  \frac{\prod_{n,\tau^{(t)}_{ln}\neq 0}\prod^{\mathcal{T}^{(t)}_{d}-1}_{u=0} \left(\alpha^{(t)}_{nk}+(\tau z)^{(t)}_{nk\backslash l}\right)}{\prod^{\mathcal{T}^{(t)}_{l}-1}_{u=0}\left[\sum^{N}_{n=1}\left(\alpha^{(t)}_{nk}+(\tau z)^{(t)}_{nk\backslash l}\right) + u \right]},
\end{align}
and
\begin{align}
  \label{eq: y Collapsed Gibbs Sampling TMDMC}
  &P(y^{(t)}_{lik}=1|\tau^{(t)},w^{(t)},z^{(t)},y^{(t)}_{\backslash li}) \nonumber \\
  &\propto 
  \left(\eta^{(t)}_{k}+z^{(t)}_{lk}+Y^{(t)}_{lk\backslash i}\right)
  \frac{\beta^{(t)}_{w_{li}}+Y^{(t)}_{w_{li}k\backslash li}}{\sum^{V}_{w=1}\left(\beta^{(t)}_{w}+Y^{(t)}_{wk\backslash li}\right)},
\end{align}
respectively.
Here, we have introduced
$(\tau z)^{(t)}_{nk\backslash l} = \sum^{L}_{l'\neq l}\tau^{(t)}_{l'n}z^{(t)}_{l'k}$,
$Y^{(t)}_{l.k}=\sum^{N_{l}}_{i=1}y^{(t)}_{lik}$,
$Y^{(t)}_{lk\backslash i}=\sum^{N_{l}}_{i'\neq i}y^{(t)}_{li'k}$,
and 
$Y^{(t)}_{wk\backslash li}=\sum_{(l',i')\neq(l,i)}\delta_{w,w_{l'i'}} y^{(t)}_{l'i'k}$.

\paragraph{Update equations.}
In this paper, we update the hyperparameters $\alpha^{(t)}_{k}$, $\beta^{(t)}_{wk}$, and $\eta^{(t)}_{k}$ at the ends of each Markov step by approximately maximizing the likelihood function with Newton's method and Minka's fixed-point iteration~\cite{Minka2000}.
First, the parameter $\alpha^{(t)}_{k}$ is updated as
\begin{align}
  \label{eq: update of alpha}
  \alpha^{(t+1)}_{k} = \alpha^{(t)}_{k} - \frac{F_{k}(\alpha^{(t)}_{k})}{F'_{k}(\alpha^{(t)}_{k})},
\end{align}
with the log-derivative of the likelihood function
$F_{k}(\alpha^{(t)}_{k})
= \frac{d}{d\alpha^{(t)}_{k}} \ln P(\tau^{(t)},w^{(t)}|z^{(t)},y^{(t)})$.
Second, the parameters $\beta^{(t+1)}_{wk}$ can be estimated with Minka's fixed-point iteration as
\begin{align}
  \label{eq: update of beta}
  &\beta^{(t+1)}_{wk}
  =  \frac{\left[ \Psi\left(\beta^{(t)}_{wk}+Y^{(t)}_{wk} \right)
  - \Psi\left(\beta^{(t)}_{wk}\right) \right] \beta^{(t)}_{wk}}
  {\left[ \Psi\left(\beta^{(t)}_{k}+Y^{(t)}_{k} \right)
  - \Psi\left(\beta^{(t)}_{k}\right) \right]} ,
\end{align}
where
$\beta^{(t)}_{k}=\sum^{V}_{w=1}\beta^{(t)}_{wk}$,
and $Y^{(t)}_{k}=\sum^{V}_{w=1} Y^{(t)}_{wk}$.
Finally, the parameter $\eta^{(t+1)}_{k}$ for the next time step is also obtained by using Minka's fixed-point iteration as
\begin{align}
  \label{eq: update of eta}
  &\eta^{(t+1)}_{k}
  =  \frac{\sum^{L}_{l=1}\left[ \Psi\left(\eta^{(t)}_{k}+Y^{(t)}_{l.k}+z^{(t)}_{lk} \right)
  - \Psi\left(\eta^{(t)}_{k}\right) \right]\eta^{(t)}_{k}}
  {\sum^{L}_{l=1}\left[ \Psi\left(\eta^{(t)}_{\rm sum}+Y^{(t)}_{l..}+1 \right)
  - \Psi\left(\eta^{(t)}_{\rm sum}\right) \right]} ,
\end{align}
with
$\eta^{(t)}_{\rm sum}=\sum^{K}_{k=1}\eta^{(t)}_{k}$
and
$Y^{(t)}_{l..} = \sum^{K}_{k=1}Y^{(t)}_{l.k}$.

\paragraph{Inference of parameters at next time step.}
We can obtain the expectation value of the probability $p^{(t)}(n|k)$ by using its posterior Dirichlet distribution as
\begin{align}
  \label{eq: p_t Collapsed Gibbs Sampling}
  p^{(t)}(n|k)
  &= \frac{\alpha^{(t)}_{nk} + \left(\tau z\right)^{(t)}_{nk}}{\sum^{N}_{n=1}\left[\alpha^{(t)}_{nk}+\left(\tau z\right)^{(t)}_{nk}\right]}.
\end{align}
The updated parameters $\alpha^{(t+1)}_{k}$ and the estimate (\ref{eq: p_t Collapsed Gibbs Sampling}) of $p^{(t)}$ are used to compute the parameters $\alpha^{(t+1)}_{nk}$ for the prior distribution of $p^{(t+1)}$ at the next time.

\paragraph{Algorithm.}

The sampling algorithms of latent variables $z^{(t)}$ and $y^{(t)}$ and the update equations of parameters $\alpha^{(t)}$, $\beta^{(t)}$, $\eta^{(t)}$, and $p^{(t)}$ are summarized in Algorithm \ref{alg: Collapsed Gibbs Sampling}.
The numbers of time steps and Monte Carlo sampling are denoted by $T_{\rm step}$ and $S$, respectively.

\begin{algorithm}[tb]
\caption{TMDMC}         
\label{alg: Collapsed Gibbs Sampling}
\begin{algorithmic}[1]
\STATE Initialize $\alpha^{(1)}$, $\beta^{(1)}$, $\eta^{(1)}$, and $p^{(0)}$
\FOR{$t=1,2,\cdots,T_{\rm step}$}
\STATE Initialize $z^{(t)}$ and $y^{(t)}$
\FOR{$s=1,2,\cdots,S$}
\FOR{$l=1,2,\cdots,L$}
\STATE Draw $z^{(t)}_{l}$ with Eq.~(\ref{eq: z Collapsed Gibbs Sampling TMDMC})
\FOR{$i=1,2,\cdots,N_{l}$}
\STATE Draw $y^{(t)}_{li}$ with Eq.~(\ref{eq: y Collapsed Gibbs Sampling TMDMC})
\ENDFOR
\ENDFOR
\ENDFOR
\STATE Update $\alpha^{(t+1)}$ with Eq.~(\ref{eq: update of alpha})
\STATE Update $\beta^{(t+1)}$ with Eq.~(\ref{eq: update of beta})
\STATE Update $\eta^{(t+1)}$ with Eq.~(\ref{eq: update of eta})
\STATE Estimate $p^{(t)}$ with Eq.~(\ref{eq: p_t Collapsed Gibbs Sampling})
\ENDFOR
\end{algorithmic}
\end{algorithm}

\subsection{Semi-supervised Node Classification}
\label{subsec: Semi-supervised}

We apply TMDMC to semi-supervised node-classification tasks, where class labels of a small part of nodes are given to predict those of the remaining nodes.
It is straightforward to generalize generative models to semi-supervised setups when class labels are translated to latent variables.
In the following experiments, we determine the latent variables from labeled data as follows:
(1) Community assignments $z^{(t)}_{l}$ of link $l$ is set as the corresponding class when either of the endpoint nodes is labeled.
When both of the endpoint nodes are labeled, randomly choose one of the class labels.
(2) Topic assignments $y^{(t)}$ of all the words in labeled documents are set as the ground-truth label.
The protocols (1) and (2) can be used for LDA-based models and original MDMC as well because they share common latent variables with TMDMC.

Community structures can be detected by TMDMC in two different ways in terms of community and topic distributions.
First, agent probability distributions $p^{(t)}$ can be used to predict community distributions of node $n$ with
$p^{(t)}_{kn} = p^{(t)}(n|k) \pi^{(t)}_{k}/ p^{(t)}_{n}$,
where $\pi^{(t)}_{k} = \eta^{(t)}_{k} / \sum^{K}_{k=1} \eta^{(t)}_{k}$
and
$ p^{(t)}_{n} = \sum^{K}_{k=1} p^{(t)}(n|k) \pi^{(t)}_{k}$.
Another way to quantify community assignments is to consider topic distributions of words contained in the corresponding documents.
When we focus on document $d$ with $N_{d}$ words, the posteriori topic distribution is given by $\theta^{(t)}_{dk} = (Y^{(t)}_{d.k} + \eta^{(t)}_{k})/ \sum^{K}_{k=1} (Y^{(t)}_{d.k} + \eta^{(t)}_{k})$ with $Y^{(t)}_{d.k}=\sum^{N_{d}}_{i=1}y^{(t)}_{dik}$.
The expectation value of $\theta^{(t)}_{dk}$ can be evaluated with Gibbs samplers.
The first indicator $p^{(t)}_{kn}$ is used in previous MDMC papers, while the second one $\theta^{(t)}_{dk}$ is useful in LDA-based models.

\section{Experiments}
\label{sec: experiments}

In this section, we evaluate the performance of TMDMC for semi-supervised node classification tasks with real-world citation networks.
We describe experimental setups and baseline methods in Subsections~\ref{subsec: Experimental Setup} and \ref{subsec: baselines}, respectively.
Classification accuracies of TMDMC are presented and compared with various methods in Subsections~\ref{subsec: result}.

\subsection{Experimental Setup}
\label{subsec: Experimental Setup}

We evaluate TMDMC with two citation networks: Citeseer and Cora~\cite{sen2008collective}.
We also analyze the largest connected components (LCCs) of the networks to study effects of connectivity.
The datasets contain BoW-represented documents and citation links between them.
Each document is classified into a single ground-truth class.
We consider sets of words in the documents as feature vectors, and construct undirected graphs from the citation links.
The statistics of these datasets are summarized in Table \ref{tab: dataset statistics}.


We conduct semi-supervised node classification experiments in the following setups:
We randomly sample a few percent of nodes for each class as labeled data, and the rest of the nodes are left for unlabeled data.
Classes of labeled nodes are translated to fixed values of latent variables with the protocol described in Subsection \ref{subsec: Semi-supervised}.
Model parameters are optimized in a transductive manner, where both labeled and unlabeled data are observed in parameter inference steps.
Predictive performance of models is evaluated with accuracy of classes assigned to unlabeled data.
We have repeated these procedures $10$ times for each dataset to perform statistical analysis with various data splits.

The model parameters in TMDMC are set as $\alpha^{(0)}_{k}= 0.1 L$, $\beta^{(0)}_{vk}=$1, $\eta^{(0)}_{k}= 1$ for all $k$ and $v$, and are updated as described in Subsection \ref{subsec: Parameter Inference}.
The number of communities $K$ and that of vocabulary $V$ are read from the ground-truth classes and the dimension of feature vectors.
Gibbs sampling is performed with sampling size $S=100$, and burn-in period $S_{\rm burn}=100$.
The maximum Markov time is set as $T_{\rm step}=20$.

\begin{table}
\centering
\setlength\tabcolsep{5pt}
\begin{tabular}{lrrrrr}  
\toprule
Dataset  & Nodes  & Edges  & Clusters & Classes & Features \\
\midrule
Citeseer & 3,312  & 4,732  & 437 &  6 & 3,703 \\
\ \ \ \ -LCC     & 2,110  & 3,757  & 1   &  6 & 3,703 \\
Cora     & 2,708  & 5,429  & 77  &  7 & 1,433 \\
\ \ \ \ -LCC & 2,485  & 5,209  & 1   &  7 & 1,433 \\
\bottomrule
\end{tabular}
\caption{Dataset statistics.}
\label{tab: dataset statistics}
\end{table}

\begin{table}[t]
\centering
\setlength\tabcolsep{4pt}
\begin{tabular}{lccll}  
\toprule
Method    & Text       & Graph    & Model      & Classifier\\
\midrule
TSVM      & \checkmark & -          & Vector-space & One-vs-rest\\
LP        & -          & \checkmark & Graph-based  & Neighbor-voting\\
Planetoid & \checkmark & \checkmark & Neural       & Softmax\\
\midrule
LDA       & \checkmark & -          & Generative   & Argmax$(p_{kd})$\\
MDMC      & -          & \checkmark & Generative   & Argmax$(p_{kn})$\\
gRTM      & \checkmark & \checkmark & Generative   & Argmax$(p_{kd})$\\
TMDMC     & \checkmark & \checkmark & Generative   & Argmax($p_{kn},p_{kd})$\\
\bottomrule
\end{tabular}
\caption{Summary of baseline and proposed methods.}
\label{tab: methods}
\end{table}

\subsection{Baselines}
\label{subsec: baselines}

We compare TMDMC against previous methods developed for transductive classification: 
transductive support vector machine (TSVM)~\cite{joachims1999transductive}, label propagation (LP)~\cite{zhu2003semi}, and Planetoid~\cite{Yang:2016:RSL:3045390.3045396}.
TSVM and LP can leverage only feature vectors and graph structures, respectively, while Planetoid utilizes both of the data.
We build one-vs-the-rest classifiers with TSVM implemented in SVMLight\footnote{\url{http://svmlight.joachims.org/}}, and use python-igraph\footnote{\url{https://igraph.org/2014/02/04/igraph-0.7-python.html}} library for LP.
For Planetoid, we use a transductive version provided by the authors\footnote{\url{https://github.com/kimiyoung/planetoid}}.
We have also implemented LDA~\cite{Blei:2003:LDA:944919.944937}, MDMC~\cite{suzuki2019bayesian}, and gRTM~\cite{Chen2015} to perform comparative studies of TMDMC with other generative models.
Labeled data are used in the same way as described above when a model has latent variables corresponding to $y$ and $z$ in TMDMC.
Similarly, the initial values and the updated equations of the hyper-parameters in LDA, MDMC, and gRTM are common in TMDMC.
We use the hinge loss in gRTM with cost parameters $l=1$ and $c=4$.
The baseline and proposed methods are summarized in Table~\ref{tab: methods}, which are divided into upper and lower rows depending on whether they are generative models or not.


\subsection{Experimental Results}
\label{subsec: result}

\begin{table}[t]
\centering
\setlength\tabcolsep{3pt}
\begin{tabular}{llll}  
\toprule
Method    & Citeseer [-LCC]          & Cora [-LCC]       \\
\midrule
TSVM      & 58.0 $\pm$ 0.8 [59.8 $\pm$ 1.1] & 50.0 $\pm$ 0.7 [52.1 $\pm$ 0.7] \\
LP        & 47.2 $\pm$ 0.8 [60.2 $\pm$ 1.7] & 62.6 $\pm$ 1.5 [63.8 $\pm$ 1.9] \\
Planetoid & \underline{60.9} $\pm$ 0.5 [\underline{66.1} $\pm$ 1.7] & 65.7 $\pm$ 0.9 [69.6 $\pm$ 0.8] \\
\midrule
LDA       & {\bf 60.0} $\pm$ 1.0 [55.7 $\pm$ 2.7] & 55.2 $\pm$ 1.0 [54.9 $\pm$ 1.6] \\
MDMC      & 34.1 $\pm$ 1.0 [40.2 $\pm$ 1.4] & 53.4 $\pm$ 1.4 [52.8 $\pm$ 2.2] \\
gRTM      & 58.6 $\pm$ 1.1 [60.0 $\pm$ 1.0] & 56.7 $\pm$ 1.0 [56.7 $\pm$ 0.5] \\
TMDMC     & 57.8 $\pm$ 1.2 [{\bf 61.0} $\pm$ 1.2] & \underline{{\bf 67.7}} $\pm$ 1.2 [\underline{{\bf 70.0}} $\pm$ 1.1] \\
\bottomrule
\end{tabular}
\caption{Classification accuracy in percent with $3\%$-labeled data.}
\label{tab: result statistics 0.03}
\end{table}

\begin{table}[t]
\centering
\setlength\tabcolsep{3pt}
\begin{tabular}{llll}  
\toprule
Method    & Citeseer [-LCC]         & Cora [-LCC]       \\
\midrule
TSVM      & 60.0 $\pm$ 0.2 [62.4 $\pm$ 0.2] & 56.2 $\pm$ 0.3 [56.4 $\pm$ 0.4] \\
LP        & 50.2 $\pm$ 0.7 [64.8 $\pm$ 0.7] & 67.4 $\pm$ 1.0 [69.0 $\pm$ 1.2] \\
Planetoid & \underline{65.9} $\pm$ 0.3 [\underline{69.1} $\pm$ 0.6] & \underline{74.7} $\pm$ 0.5 [\underline{76.0} $\pm$ 0.6] \\
\midrule
LDA       & {\bf 61.9} $\pm$ 0.6 [61.9 $\pm$ 0.4] & 59.0 $\pm$ 0.7 [58.6 $\pm$ 0.6] \\
MDMC      & 38.7 $\pm$ 0.6 [44.7 $\pm$ 2.0] & 62.4 $\pm$ 0.6 [62.8 $\pm$ 1.7] \\
gRTM      & 59.9 $\pm$ 1.1 [62.3 $\pm$ 0.4] & 58.2 $\pm$ 0.6 [60.3 $\pm$ 0.5] \\
TMDMC     & 59.7 $\pm$ 0.9 [{\bf 62.7} $\pm$ 0.7] & {\bf 71.0} $\pm$ 0.5 [{\bf 72.3} $\pm$ 0.4] \\
\bottomrule
\end{tabular}
\caption{Classification accuracy in percent with $5\%$-labeled data.}
\label{tab: result statistics 0.05}
\end{table}

\begin{figure*}[t]
\centering
 \begin{minipage}{0.45\hsize}
  \begin{center}
   \includegraphics[width=80mm]{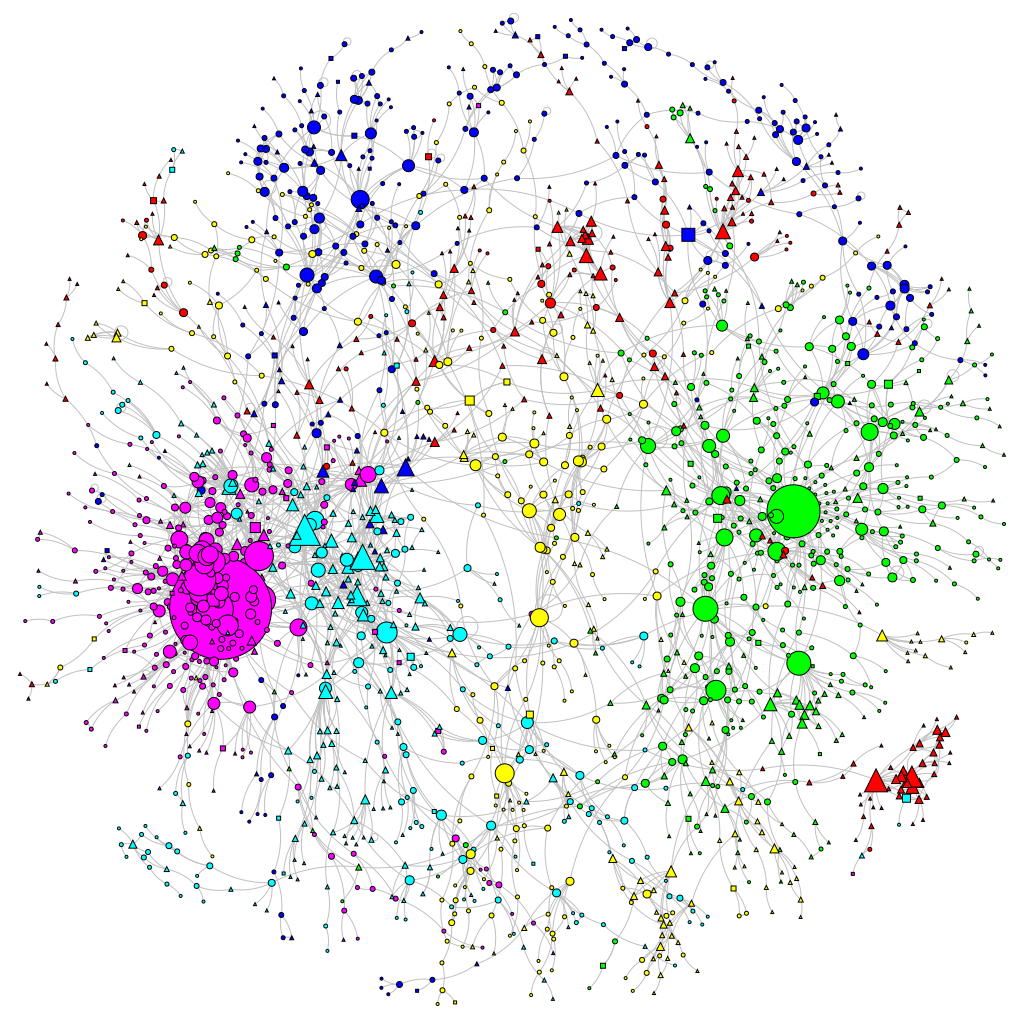}
  \end{center}
  \caption{Communities of Citeseer-LCC with TMDMC.}
  \label{fig: citeseer}
 \end{minipage}
 \begin{minipage}{0.45\hsize}
  \begin{center}
   \includegraphics[width=80mm]{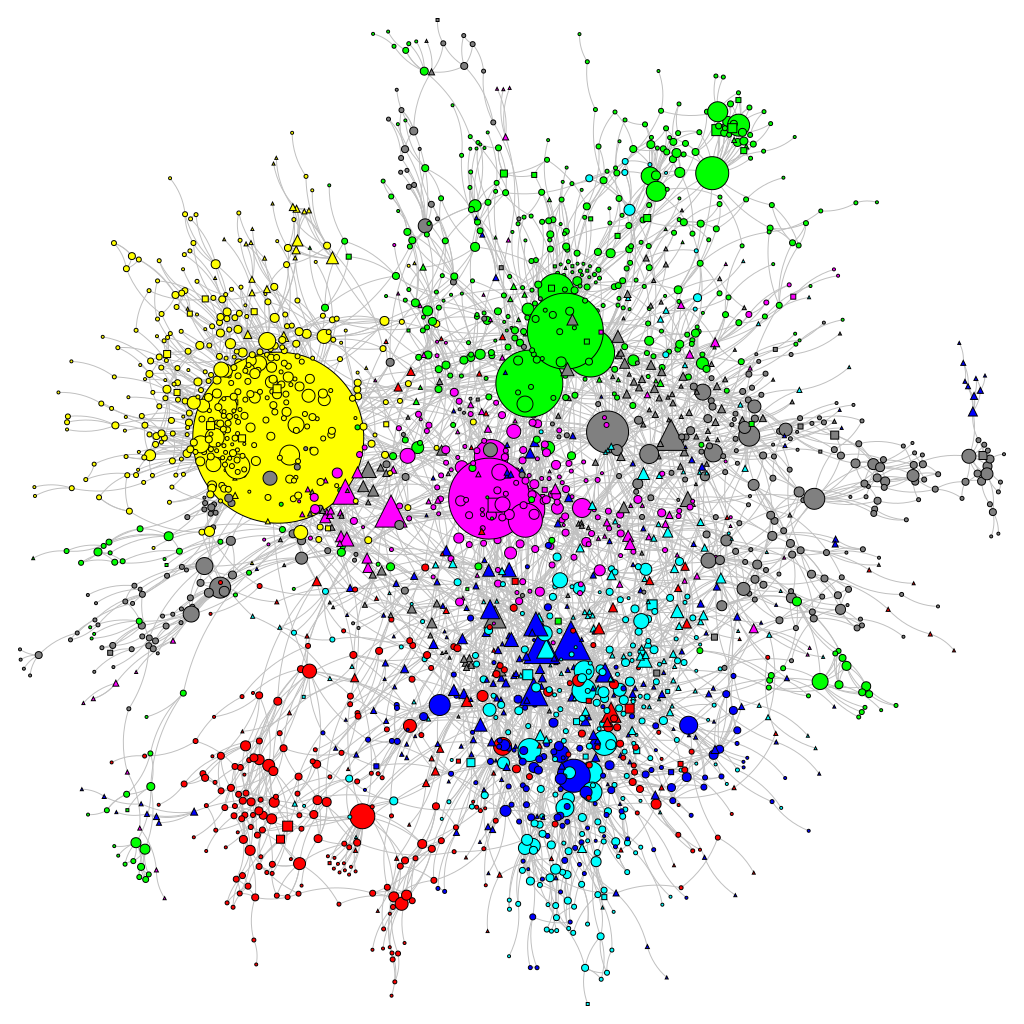}
  \end{center}
  \caption{Communities of Cora-LCC with TMDMC.}
  \label{fig: cora}
 \end{minipage}
\end{figure*}

Table~\ref{tab: result statistics 0.03} shows means and standard errors of classification accuracies in $3\%$-labeled datasets out of $10$ trials.
Values in square brackets are results of LCCs in corresponding datasets.
We make values underscored (bold) when it is highest among all (generative) models.
In terms of expectation values, TMDMC outperforms other generative models, i.e. LDA, MDMC, and gRTM, in Citeer-LCC, Cora and Cora-LCC.
In addition, TMDMC achieves the best score in Cora datasets including non-generative models.
Citeseer is characteristic in that feature-based methods clearly outperforms genuinely graph-based methods.
Besides, the performance of LP, Planetoid, MDMC, and TMDMC, all of which are based on random walks, is significantly improved when LCC is considered.
This indicates that observed graph structures are not so consistent with ground-truth labels, while texts are informative for classification.
In contrast to Citeseer, both feature- and graph-based methods perform evenly well in Cora.
It is also worth noting that TMDMC surpasses Planetoid, which employs deep neural architectures, in Cora and Cora-LCC.
These observations imply that TMDMC can effectively learn community patterns of document networks from limited numbers of labeled data especially
when both texts and graphs can collaboratively predict ground-truth class labels.

In order to evaluate the size effect, we test the models with increased proportions of the labeled data.
Table~\ref{tab: result statistics 0.05} reports classification accuracies with $5\%$ labeled data.
While TMDMC still performs better than other generative models in Citeseer-LCC, Cora, and Cora-LCC, Planetoid wins the highest scores in all the datasets.
In particular, the margin between Planetoid and TMDMC gets wider in Citeseer.
We conjecture that high-capacity neural architectures employed by Planetoid could discern informative and uninformative features with increased training data.
Discernibility is crucial especially in Citeseer datasets because observed graph may sometimes cause confusion in predicting correct class labels.





We finally visualize typical results of $3\%$-labeled Citeseer-LCC (Cora-LCC) in Figure~\ref{fig: citeseer} (\ref{fig: cora}) to qualitatively analyze characteristics of TMDMC.
Node sizes are scaled to their degrees, and node colors represent class labels assigned by TMDMC.
Layouts are computed with the Fruchterman–Reingold algorithm.
Circle, triangle and square nodes denote correctly-classified, misclassified and labeled nodes, respectively.
As expected, nodes with large degrees tend to be correctly classified by TMDMC, where random walkers make account of central nodes in link generation processes.
Besides, nodes of the same label tend to aggregate, even in misclassified cases, into clusters, because of the associative natures of TMDMC modeling.
The second feature results in misclassification in Citeseer, where peripheral nodes do not necessarily belong to the same community as central ones.
In Cora, TMDMC fails to distinguish ``blue'' and ``cyan'' communities, which are densely connected in networks and require classifiers to find additional discriminating features.








\section{Conclusions and Future Works}
\label{sec: discussion and conclusion}

In this paper, we have proposed a novel generative model named TMDMC for analyzing document networks.
The proposed model unifies random-walk-based community detection and topic modeling to jointly model graph structures and textual information.
Random walkers quantify node centrality, which is incorporated in link generation processes.
We have compared TMDMC with previous probabilistic models, i.e. LDA, MDMC, and gRTM, in semi-supervised classification tasks.
TMDMC outperforms other probabilistic models in classifying nodes in connected components of real-world citation networks.
Besides, TMDMC surpasses Planetoid, a deep-neural model, in $3\%$-labeled Cora dataset.
This indicates that TMDMC detects community structures from a limited number of labeled data.
For future works, it is promising to enhance model capacities of TMDMC with neural architectures to discern informative data in a similar way to NRTM \cite{Bai2018}.
This extension can strengthen modeling of mutual interactions between text and graph natures as well.




\section*{Acknowledgments}
The author would like to thank Xule Qiu, and Seiya Inagi for fruitful discussions and comments.

\bibliographystyle{named}
\bibliography{paper}

\begin{thebibliography}{}

\bibitem[\protect\citeauthoryear{Bai \bgroup \em et al.\egroup
  }{2018}]{Bai2018}
Haoli Bai, Zhuangbin Chen, Michael~R. Lyu, Irwin King, and Zenglin Xu.
\newblock {Neural relational topic models for scientific article analysis}.
\newblock {\em International Conference on Information and Knowledge Management
  (CIKM'18)}, pages 27--36, 2018.

\bibitem[\protect\citeauthoryear{Belkin \bgroup \em et al.\egroup
  }{2006}]{belkin2006manifold}
Mikhail Belkin, Partha Niyogi, and Vikas Sindhwani.
\newblock Manifold regularization: A geometric framework for learning from
  labeled and unlabeled examples.
\newblock {\em Journal of Machine Learning Research}, 7:2399--2434, 2006.

\bibitem[\protect\citeauthoryear{Blei \bgroup \em et al.\egroup
  }{2003}]{Blei:2003:LDA:944919.944937}
David~M. Blei, Andrew~Y. Ng, and Michael~I. Jordan.
\newblock Latent dirichlet allocation.
\newblock {\em Journal of Machine Learning Research}, 3:993--1022, 2003.

\bibitem[\protect\citeauthoryear{Brin and Page}{1998}]{brin1998anatomy}
Sergey Brin and Lawrence Page.
\newblock The anatomy of a large-scale hypertextual web search engine.
\newblock {\em Computer networks and ISDN systems}, 30(1-7):107--117, 1998.

\bibitem[\protect\citeauthoryear{Chang and Blei}{2009}]{pmlr-v5-chang09a}
Jonathan Chang and David Blei.
\newblock Relational topic models for document networks.
\newblock In {\em Proceedings of the 12th International Conference on
  Artificial Intelligence and Statistics}, volume~5 of {\em Proceedings of
  Machine Learning Research}, pages 81--88, 2009.

\bibitem[\protect\citeauthoryear{Chen \bgroup \em et al.\egroup
  }{2015}]{Chen2015}
Ning Chen, Jun Zhu, Fei Xia, and Bo~Zhang.
\newblock {Discriminative relational topic models}.
\newblock {\em IEEE Transactions on Pattern Analysis and Machine Intelligence},
  37(5):973--986, 2015.

\bibitem[\protect\citeauthoryear{Cohn and Hofmann}{2001}]{cohn2001missing}
David~A Cohn and Thomas Hofmann.
\newblock The missing link-a probabilistic model of document content and
  hypertext connectivity.
\newblock In {\em Advances in Neural Information Processing Systems (NIPS'00)},
  pages 430--436, 2001.

\bibitem[\protect\citeauthoryear{Erosheva \bgroup \em et al.\egroup
  }{2004}]{Erosheva5220}
Elena Erosheva, Stephen Fienberg, and John Lafferty.
\newblock Mixed-membership models of scientific publications.
\newblock {\em Proceedings of the National Academy of Sciences},
  101:5220--5227, 2004.

\bibitem[\protect\citeauthoryear{Fortunato and Hric}{2016}]{FORTUNATO20161}
Santo Fortunato and Darko Hric.
\newblock Community detection in networks: A user guide.
\newblock {\em Physics Reports}, 659:1 -- 44, 2016.

\bibitem[\protect\citeauthoryear{Joachims}{1999}]{joachims1999transductive}
Thorsten Joachims.
\newblock Transductive inference for text classification using support vector
  machines.
\newblock In {\em Proceedings of the 16th International Conference on Machine
  Learning (ICML'99)}, pages 200--209, 1999.

\bibitem[\protect\citeauthoryear{Kipf and Welling}{2017}]{Kipf2017}
Thomas~N. Kipf and Max Welling.
\newblock Semi-supervised classification with graph convolutional networks.
\newblock In {\em International Conference on Learning Representations
  (ICLR'17)}, 2017.

\bibitem[\protect\citeauthoryear{Liu \bgroup \em et al.\egroup
  }{2009}]{Liu2009}
Yan Liu, Alexandru Niculescu-Mizil, and Wojciech Gryc.
\newblock {Topic-link LDA: Joint models of topic and author community}.
\newblock {\em Proceedings of the 26th International Conference On Machine
  Learning (ICML'09)}, pages 665--672, 2009.

\bibitem[\protect\citeauthoryear{Minka}{2000}]{Minka2000}
Thomas~P. Minka.
\newblock {Estimating a Dirichlet distribution}.
\newblock \url{https://tminka.github.io/papers/dirichlet/minka-dirichlet.pdf},
  2000.

\bibitem[\protect\citeauthoryear{Nallapati and Cohen}{2008}]{nallapati2008link}
Ramesh Nallapati and William~W Cohen.
\newblock Link-plsa-lda: A new unsupervised model for topics and influence of
  blogs.
\newblock In {\em International Conference on Weblogs and Social Media
  (ICWSM'08)}, pages 84--92, 2008.

\bibitem[\protect\citeauthoryear{Okamoto and Qiu}{2018}]{mdmc_original}
Hiroshi Okamoto and Xule Qiu.
\newblock Community detection by modular decomposition of random walk.
\newblock In {\em The 7th International Conference on Complex Networks and
  Their Applications}, page~59, 2018.

\bibitem[\protect\citeauthoryear{Okamoto and Qiu}{2019}]{okamoto2019modular}
Hiroshi Okamoto and Xule Qiu.
\newblock Modular decomposition of markov chain: detecting hierarchical
  organization of pervasive communities.
\newblock {\em arXiv preprint}, arXiv:1909.07066, 2019.

\bibitem[\protect\citeauthoryear{Perozzi \bgroup \em et al.\egroup
  }{2014}]{perozzi2014deepwalk}
Bryan Perozzi, Rami Al-Rfou, and Steven Skiena.
\newblock Deepwalk: Online learning of social representations.
\newblock In {\em Proceedings of the 20th ACM SIGKDD International Conference
  on Knowledge Discovery and Data Mining (KDD'14)}, pages 701--710, 2014.

\bibitem[\protect\citeauthoryear{Sen \bgroup \em et al.\egroup
  }{2008}]{sen2008collective}
Prithviraj Sen, Galileo Namata, Mustafa Bilgic, Lise Getoor, Brian Galligher,
  and Tina Eliassi-Rad.
\newblock Collective classification in network data.
\newblock {\em AI magazine}, 29:93--93, 2008.

\bibitem[\protect\citeauthoryear{Suzuki}{2019}]{suzuki2019bayesian}
Takafumi~J. Suzuki.
\newblock Bayesian modeling of random walker for community detection in
  networks.
\newblock {\em arXiv preprint}, arXiv:1910.11587, 2019.

\bibitem[\protect\citeauthoryear{Terragni \bgroup \em et al.\egroup
  }{2020}]{TERRAGNI2020581}
Silvia Terragni, Elisabetta Fersini, and Enza Messina.
\newblock Constrained relational topic models.
\newblock {\em Information Sciences}, 512:581--594, 2020.

\bibitem[\protect\citeauthoryear{Weston \bgroup \em et al.\egroup
  }{2012}]{weston2012deep}
Jason Weston, Fr{\'e}d{\'e}ric Ratle, Hossein Mobahi, and Ronan Collobert.
\newblock Deep learning via semi-supervised embedding.
\newblock In {\em Neural Networks: Tricks of the Trade}, pages 639--655. 2012.

\bibitem[\protect\citeauthoryear{Xu \bgroup \em et al.\egroup
  }{2019}]{xu2019graph}
Bingbing Xu, Huawei Shen, Qi~Cao, Keting Cen, and Xueqi Cheng.
\newblock Graph convolutional networks using heat kernel for semi-supervised
  learning.
\newblock In {\em Proceedings of the 28th International Joint Conference on
  Artificial Intelligence (IJCAI'19)}, pages 1928--1934. AAAI Press, 2019.

\bibitem[\protect\citeauthoryear{Yang \bgroup \em et al.\egroup
  }{2016a}]{yang-etal-2016-discriminative}
Weiwei Yang, Jordan Boyd-Graber, and Philip Resnik.
\newblock A discriminative topic model using document network structure.
\newblock In {\em Proceedings of the 54th Annual Meeting of the Association for
  Computational Linguistics (ACL'16)}, pages 686--696, 2016.

\bibitem[\protect\citeauthoryear{Yang \bgroup \em et al.\egroup
  }{2016b}]{Yang:2016:RSL:3045390.3045396}
Zhilin Yang, William~W. Cohen, and Ruslan Salakhutdinov.
\newblock Revisiting semi-supervised learning with graph embeddings.
\newblock In {\em Proceedings of the 33rd International Conference on
  International Conference on Machine Learning (ICML'16)}, pages 40--48, 2016.

\bibitem[\protect\citeauthoryear{Zhu \bgroup \em et al.\egroup
  }{2003}]{zhu2003semi}
Xiaojin Zhu, Zoubin Ghahramani, and John~D Lafferty.
\newblock Semi-supervised learning using gaussian fields and harmonic
  functions.
\newblock In {\em Proceedings of the 20th International conference on Machine
  learning (ICML'03)}, pages 912--919, 2003.

\end{thebibliography}

\end{document}